\documentclass{article}

\usepackage{nac_preprint}
\usepackage[utf8]{inputenc} % allow utf-8 input
\usepackage[T1]{fontenc}    % use 8-bit T1 fonts
\usepackage{hyperref}       % hyperlinks
\usepackage{url}            % simple URL typesetting
\usepackage{booktabs}       % professional-quality tables
\usepackage{amsfonts}       % blackboard math symbols
\usepackage{nicefrac}       % compact symbols for 1/2, etc.
\usepackage{microtype}      % microtypography

\usepackage{booktabs} % For formal tables
\usepackage{subcaption}
\usepackage{amsmath}
\usepackage{mathtools}
\usepackage[toc,page]{appendix}
\usepackage{enumitem}
\usepackage{graphics}
\usepackage{graphicx}
\usepackage{xcolor}
\usepackage[export]{adjustbox}
\usepackage{algorithm}
\usepackage{algorithmicx}
\usepackage[noend]{algpseudocode}
\algnewcommand{\LineComment}[1]{\State \(//\) #1}
\algnewcommand{\RLineComment}[1]{\State \(\triangleright\) #1}
\usepackage{bm}
\usepackage{multirow}

\usepackage{url,hyperref,lineno,microtype,subcaption}
\usepackage[onehalfspacing]{setspace}
\usepackage{soul,xcolor}
\usepackage{caption, subcaption}
\usepackage{algpseudocode,amsmath}
\DeclareMathOperator*{\argmax}{\arg\max}

\DeclareUnicodeCharacter{1E6B}{${\dot \atop \tau}$}

\title{A Neural Active Inference Model of Perceptual-Motor Learning}

\author{%
  Zhizhuo Yang\\
  Computer Science Department \\
  Rochester Institute of Technology \\
  Rochester, NY 14623\\
  \texttt{zy8981@rit.edu}
  \And
  Gabriel J. Diaz\\
  Chester F. Carlson Center for Imaging Science\\
  Rochester Institute of Technology \\
  Rochester, NY 14623\\
  \texttt{gabriel.diaz@rit.edu}
  \And
  Brett R. Fajen\\
  Cognitive Science Department \\
  Rensselaer Polytechnic Institute\\
  Troy, NY, USA\\
  \texttt{fajenb@rpi.edu }
  \And
  Reynold Bailey \\
  Computer Science Department \\
  Rochester Institute of Technology \\
  Rochester, NY 14623\\
  \texttt{rjb@cs.rit.edu}
  \And
  Alexander Ororbia \\
  Computer Science Department \\
  Rochester Institute of Technology \\
  Rochester, NY 14623\\
  \texttt{ago@cs.rit.edu}
}

\begin{document}

\setlength{\abovedisplayskip}{0.065cm}
\setlength{\belowdisplayskip}{0pt}

\maketitle

\begin{abstract}
The active inference framework (AIF) is a promising new computational framework grounded in contemporary neuroscience that can produce human-like behavior through reward-based learning. In this study, we test the ability for the AIF to capture the role of anticipation in the visual guidance of action in humans through the systematic investigation of a visual-motor task that has been well-explored -- that of intercepting a target moving over a ground plane. Previous research demonstrated that humans performing this task resorted to anticipatory changes in speed intended to compensate for semi-predictable changes in target speed later in the approach. To capture this behavior, our proposed ``neural'' AIF agent uses artificial neural networks to select actions on the basis of a very short term prediction of the information about the task environment that these actions would reveal along with a long-term estimate of the resulting cumulative expected free energy. Systematic variation revealed that anticipatory behavior emerged only when required by limitations on the agent's movement capabilities, and only when the agent was able to estimate accumulated free energy over sufficiently long durations into the future. In addition, we present a novel formulation of the prior function that maps a multi-dimensional world-state to a uni-dimensional distribution of free-energy. Together, these results demonstrate the use of AIF as a plausible model of anticipatory visually guided behavior in humans. 

\end{abstract}

\section{Introduction}
\label{sec:intro}

The active inference framework (AIF)~\cite{friston2009reinforcement} is an emerging theory of neural encoding and processing that captures a wide range of cognitive, perceptual, and motor phenomena, while also offering a neurobiologically plausible means of conducting reward-based learning through the capacity to predict sensory information. The behavior of an AIF agent involves the selection of action-plans that span into the near future 
and centers around the learning of a probabilistic generative model of the world through interaction with the environment. Ultimately, the agent must take action such that it is making progress towards its goals (goal-seeking behavior) while also balancing the drive to explore and understand its environment (information maximizing behavior), adjusting the internal states of its world  to better account for the evidence that it acquires over time. As a result, AIF unifies perception, action, and learning by framing them as processes that  result from approximate Bayesian inference.

The AIF framework has been used to study a variety of reinforcement learning (RL) tasks, including the inverted pendulum problem (\emph{CartPole})~\cite{millidge2020deep, shin2021prior}, the mountain car problem (\emph{MountainCar})~\cite{friston2009reinforcement, ueltzhoffer2018deep, tschantz2020scaling, ccatal2020learning, shin2021prior} and the frozen lake problem (\emph{Frozen Lake})~\cite{sajid2021active}. Each task places different demands on motor and cognitive abilities. For instance, \emph{CartPole} requires online control of a paddle to balance a pole upright, whereas \emph{MountainCar} requires intelligent exploration of the task environment; a simple ``greedy'' policy (typical of many modern-day RL approaches) would fail to solve the problem. The popular \emph{Frozen Lake} requires skills related to spatial navigation and planning if the agent is to find the goal while avoiding unsafe states. 

One fundamental aspect of human and animal behavior that has so far not been sufficiently studied from an active inference perspective is the on-line visual guidance of locomotion.  On-line visual guidance comprises a class of ecologically important behaviors for which movements of the body are continuously regulated based on currently available visual information seen from the first-person perspective.  Some of the most extensively studied tasks include steering toward a goal \cite{warren_behavioral_2010}, negotiating complex terrain on foot \cite{diaz_pickup_2018,matthis_humans_2013}, intercepting moving targets \cite{fajen_behavioral_2007}, braking to avoid a collision \cite{fajen_learning_2006,yilmaz1995visual}, and intercepting a fly ball \cite{chapman1968catching,fajen2008reconsidering}.  For each of these tasks, researchers have formulated control strategies that capture the coupling of visual information and action.

\begin{figure}[tb]
    \centering
    \includegraphics[width=0.25\linewidth]{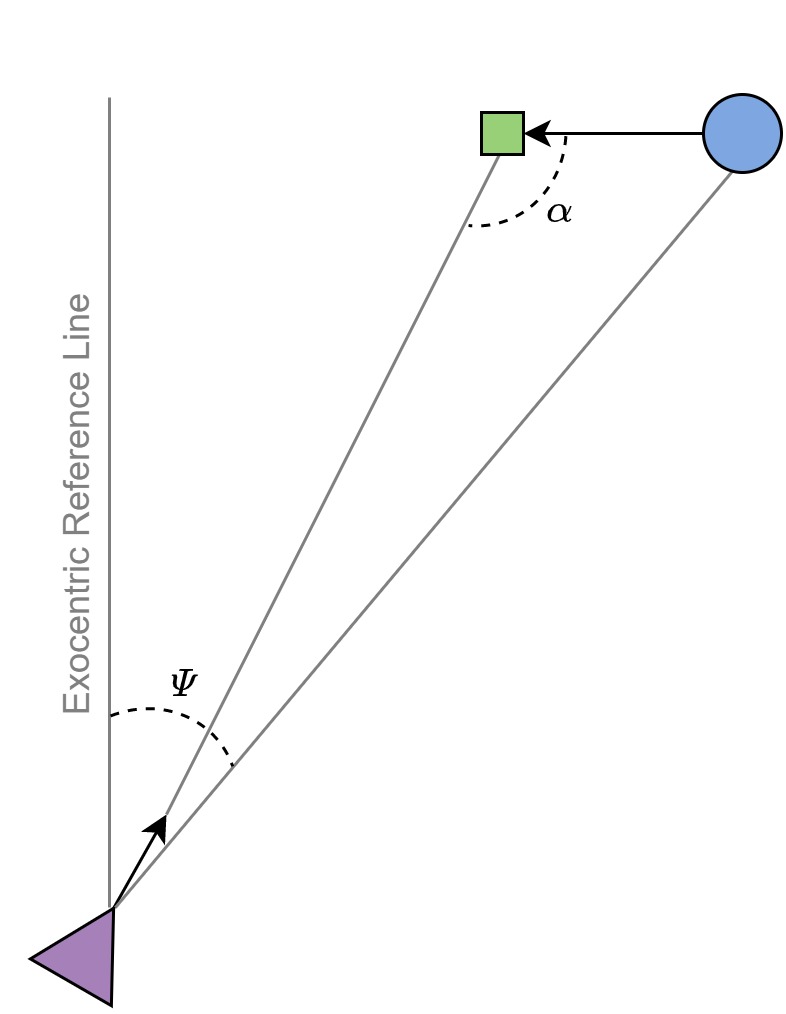}
    \caption{A top-down view of the interception problem. The
    agent (triangle) and target (circle) approach the invisible interception
    point (square) by going straight ahead. $\Psi$ denotes the exocentric direction of the target (bearing angle) and $\alpha$ denotes the target’s approach angle. Image adapted from~\cite{diaz2009intercepting}.}
    \label{fig:interception}
    \vspace{-0.5cm}
\end{figure}

One aspect of on-line visual guidance that AIF might be particularly well-suited to capture is anticipation.  To successfully perform these kinds of tasks, actors must be able to regulate their actions in anticipation of future events. One approach to capturing anticipation in visual guidance is to identify sources of visual information that specify how the actor should move at the current instant in order to reach the goal in the future. For example, when running to intercept a moving target, the sufficiency of the interceptor's current speed is specified by the rate of change in the exocentric visual direction of the target, or \textit{bearing angle} (Figure \ref{fig:interception}). If the interceptor is able to move so as to maintain a constant bearing angle (CBA), then an interception is guaranteed. Such accounts of anticipation are appealing because they avoid the need for planning on the basis of predictions or extrapolations of the agent's or target's motion, thereby presumably requiring fewer cognitive resources for task execution. Similar accounts of anticipation in the context of locomotor control have been developed for fly ball catching \cite{chapman1968catching} and braking \cite{lee_theory_1976}.

However, there are other aspects of anticipatory control that are more difficult to capture based on currently available information alone.  For example, moving targets sometimes change speeds and directions in ways that are somewhat predictable, allowing actors to alter their movement in advance in anticipation of the most likely change in target motion.  This was demonstrated in a previously published study in which subjects were instructed to adjust their self-motion speed while moving along a linear path in order to intercept a moving target that changed speed partway through each trial \cite{diaz2009intercepting}. The final target speed randomly varied between trials such that the target usually accelerated but occasionally decelerated.  In response, subjects quickly learned to adjust their speed during the first part of the trial in anticipation of the change in target speed that was most likely given past experience and the initial conditions of that trial.  

Active inference offers a potentially useful framework for understanding and modeling this kind of anticipatory behavior. The behavior of an AIF agent involves the selection of action plans (or policies) that span into the near future.  These plans are selected based on \textit{expected free energy} (EFE), i.e., a signal that takes into account both the action's contribution to reaching a desired goal state (i.e., an \textit{instrumental} component), and the new information gained by the action (i.e., an \textit{epistemic} component). This method of action selection is ideal for the study of predictive and anticipatory behavior in that it allows for the selection of action plans that do not immediately contribute to task completion, but that reveal to the agent something previously unknown about how the agent's action affects the environment. Similarly, in the task presented in~\cite{diaz2009intercepting}, the human participants learned that success required increasing speed early in the trial in order to increase the likelihood of an interception after the target's semi-predictable change in speed. Critically, this early change in speed was not motivated by currently available visual information, but rather by the positive reinforcement of actions selected in the process of task exploration.

In contrast to reinforcement learning methods, active inference (AIF) formulates action-driven learning and inference from a Bayesian, belief-based perspective~\cite{sajid2021active, parr2019generalised}. Generally, AIF offers: 
1) flexibility to define a prior preference (or preferred outcome) over the observation space (which pushes the agent to uncover goal-orienting policies), which provides an alternative to designing a reward function, 
2) a principled treatment for epistemic exploration as a means of uncertainty reduction, information gain, and intrinsic motivation~\cite{parr2017uncertainty, schwartenbeck2019computational, parr2019generalised}, and 
3) an encompassing uncertainty or precision over the beliefs that the generative model of the AIF agent computes as a natural part of then agent's belief updating~\cite{parr2017uncertainty}. Despite being a popular and powerful framework of perception, action~\cite{friston2009free, friston2010free, friston2017active, buckley2017free}, decision-making and planning~\cite{kaplan2018planning, parr2018anatomy} with biological plausibility, AIF has been mostly applied to problems with a low-dimensionality and often discrete state space and actions~\cite{friston2015active, friston2017active, firston2017curiosity, FRISTON2018486, friston2012agency}. We refer to~\cite{da2020active} for a comprehensive review on AIF.

The present study makes several contributions to the understanding of visually guided action and active inference:
\begin{itemize}[noitemsep,nolistsep]
    \item We present a novel model for locomotor interception of a target that changes speeds semi-predictably, as in~\cite{diaz2009intercepting}. This model is a generalization of AIF where EFE is treated as a negative value function in reinforcement learning (RL)~\cite{shin2021prior} and deep RL methodology is utilized to scale AIF to solve tasks such as locomotor interception with continuous state spaces. Specifically, our method predicts action-conditioned EFE values with a \textit{recognition} network (see~\ref{sec:recognition_model}) and by bootstrapping on the continuous observation space over a long time horizon. This allows the agent to  account for the long-term effects of its current chosen action(s).
    
    \item To calculate the \textit{instrumental} value, we designed a problem-specific prior function to convert the original observations into a one-dimensional prior space where a prior preference can be (more easily) specified. This allows us to inject domain knowledge into the \textit{instrumental} reward. The \textit{instrumental} measurements in prior space simultaneously promote  interpretability as well as computationally efficient task performance.
    
    \item We present a comparison of task performance of a baseline deep-Q network (DQN) agent, or an AIF agent in which EFE is computed using only the \textit{instrumental} signal/component, with a full AIF agent in which EFE is computed using both \textit{instrumental} and \textit{epistemic} signals/components.
    
    \item We demonstrate behavioral differences among our full AIF agent under the influence of two varying parameters: the discount factor $\gamma$, which describes the weight on future accumulated quantities when calculating EFE value at each time step, and pedal lag coefficient $\textit{K}$, which specifies how responsive changes in pedal position is reflected on agent's speed (or the amount of inertia that is associated with the agent's vehicle).
    
    \item We interpret our findings in as a model for anticipation in the context of visually guided action as well as in terms of specific contributions to the active inference and machine learning communities.
\end{itemize}

\section{Materials and Methods}
\label{sec:methodology}

Our aim in this study was to develop an agent that selects from a set of discrete actions in order to perform the task of interception. In this section, we describe the task that we aim to solve as well as formally describe the AIF model designed to tackle it. We start with the problem formulation and brief notation and definitions, then move on to describe our proposed agent's inference and learning dynamics. 

\subsection{The Perceptual-Motor Problem: Intercepting a Moving Target} % problem definition
\label{sec:problem_definition}

We designed and simulated a perception-motor problem based on the human interception task used by \cite{diaz2009intercepting}. In the original study, subjects sat in front of a large rear-projection screen depicting an open field with a heavily textured ground plane.  The subject's task was to intercept a moving spherical target by controlling the speed of self-movement along a linear trajectory with a foot pedal, the position of which was mapped onto speed according to a first-order lag. Subjects began each trial from a stationary position at an initial distance sampled uniformly from between $25$ and $30$ meters (m) from the interception point. The spherical target approached the subject’s path at one of three initial speeds ($11.25$, $9.47$, $8.18$ meters/second or m/s). Between $2.5$ and $3.25$ s after the trial began, the target changed speeds linearly by an amount that was sampled from a normal distribution of possible final speeds. The mean of the distribution was $15$ m/s such that target speed usually increased, but occasionally decreased (standard deviation was $5$ m/s, final speed is truncated by one standard deviation from the mean). The change of target speed takes exactly $500$ ms.

In \cite{diaz2009intercepting}, subjects were found to increase their speed during the early part of the trial in order to anticipate the most likely change in target speed, which helped them perform at near optimal levels. Differences between the behavior of subjects and the ideal pursuer were also found under some conditions. Findings in the original study further yielded insight into the strategies that humans adopt when dealing with uncertainty in realistic interception tasks.

\subsection{Notation}
\label{sec:notation}

We next define the notation and mathematical operators that we will use throughout the rest of this paper. $\odot$ indicates a Hadamard product, $\cdot$ indicates a matrix/vector multiplication (or dot product if the two objects it is applied to are vectors of the same shape), and $(\mathbf{v})^T$ denotes the transpose. $||\mathbf{v}||_p$ is used to represent the $p$-norm where $p = 2$ results in the $2$-norm or Euclidean (L2) distance.
%$|\Theta|$ denotes the cardinality of the set $\Theta$. 
%$\mathbf{M}[:,i]$ is the (column) slice operator, which means that we extract the $i$th column vector of matrix $\mathbf{M}$, whereas  $\mathbf{M}[i,:]$ is the row slice operator, where we extract the $i$th row.

\subsection{Action and Input Space Specification}
\label{sec:space_descriptions}

To simplify the problem for this work, we assume that the mapping between environmental (latent) states and observations is the identity matrix. Furthermore, we formulate the problem as a Markov Decision Process (MDP) with a discrete action space. The action space $\mathbf{a}_t$ (action vector at time $t$) is defined as a one-hot vector $\mathbf{a} \in \{0,1\}^{6 \times 1}$, where each dimension corresponds to a unique action and the actions are mutually exclusive. Each dimension corresponds to one of the pedal speeds (m/s) in $\{2, 4, 8, 10, 12, 14\}$ respectively. Once a pedal speed is selected, the agent will change its own speed by the amount of $\Delta V=\textit{K}*(V_p-V_s)$ in one time step where $V_p$ is pedal speed, $V_s$ is current subject speed and $\textit{K}$ is the lag coefficient.
Similar to Tschantz et.al~\cite{tschantz2020learning}, we assume that the control state vector (which, in AIF, control states are originally treated separately from action states) lines up one-to-one with the action vector, meaning that it too is a vector of the form $\mathbf{u} \in \{0,1\}^{6 \times 1}$. We define the observation/state space ($\mathbf{o} \in \mathcal{R}^{4 \times 1}$) to be a 4-dimensional vector $\mathbf{o}_t = \langle x_t, v_t, x_s, v_s\rangle ^T$, which corresponds to target distance, target speed, subject distance and subject speed. All distances aforementioned are with respect to the invisible interception point.

\begin{figure}[tb]
    \centering
    \includegraphics[width=0.8\textwidth]{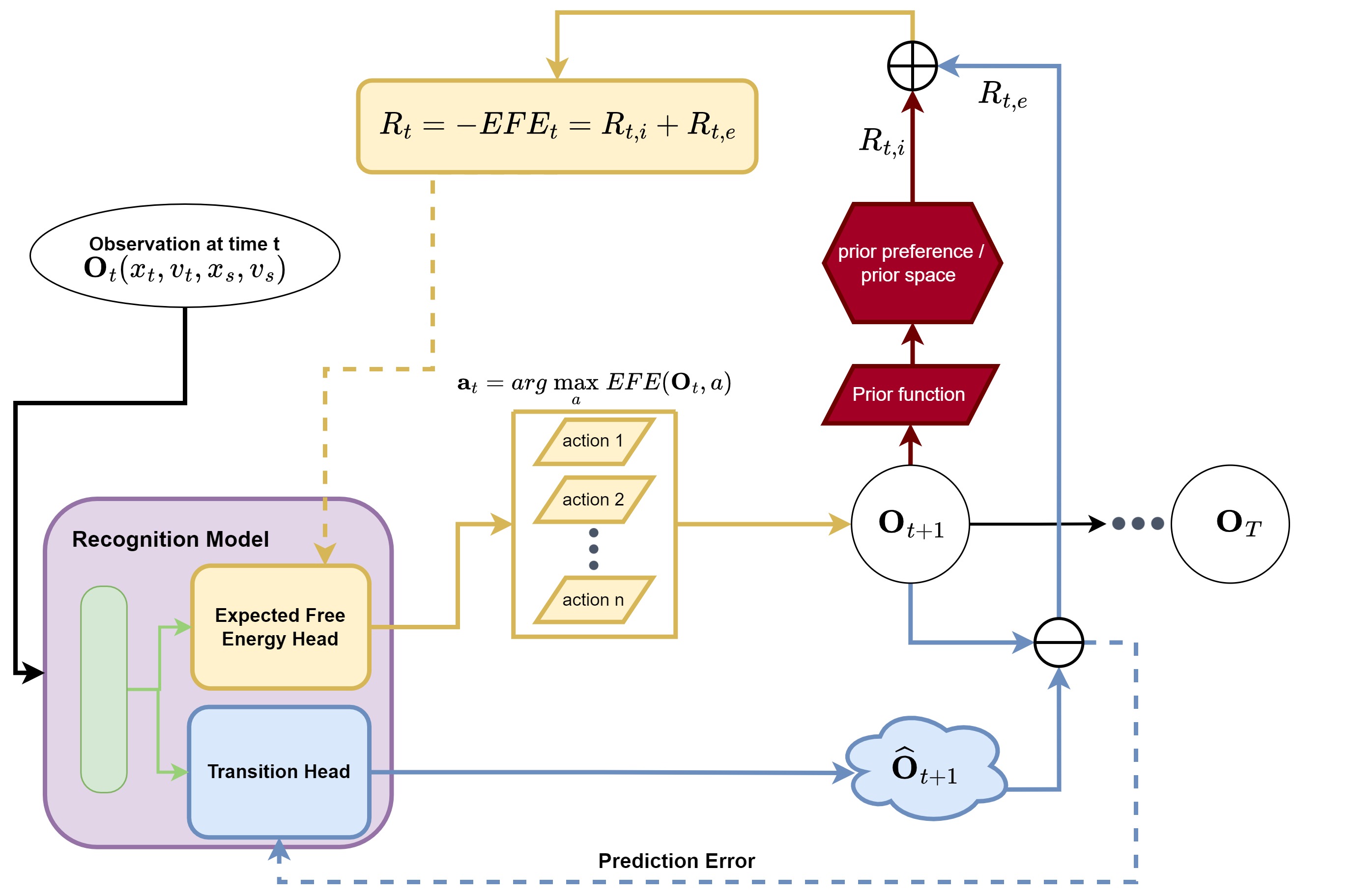}
    \caption{Our neural AIF architecture for the interception task. The recognition model is a two-headed artificial neural network which consists of shared hidden layers, an EFE (estimation) head, and a transition dynamics (prediction) head. The EFE head estimates EFE values for all possible actions given the current (latent/hidden) state. An action which is associated with maximum EFE value is selected and executed in the environment and the resulting observation is fed into the prior function where the \textit{instrumental} value $R_{t,i}$ is calculated in prior space. Meanwhile, the transition dynamics head predicts the resulting observation given the current (latent/hidden) state. The error between the predicted and actual observation at $t+1$ forms the \textit{epistemic} value $R_{t,e}$. The summation of $R_{t,i}$ and $R_{t,e}$ results in the final EFE (target) value.}
    \label{fig:AIF_arch}
    \vspace{-0.5cm}
\end{figure}

\subsection{Neural Active Inference}
\label{sec:aif_arch}

% Describe G-learning, the instrumental term and the epistemic term
Active inference (AIF) is a (Bayesian) computational framework that brings together perception and action under one single imperative: minimizing \emph{free energy}. It accounts for how self-organizing agents operate in dynamic, non-stationary environments \cite{friston2019free}, offering an alternative to standard, reward function-centric reinforcement learning (RL). 
In this study, we craft a simple AIF agent that resembles Q-learning~\cite{shin2021prior} where the \textit{expected free energy} (EFE) serves the role of a negative action-value function in RL.
We frame the definition of EFE in the context of a stochastic policy and cast action-conditioned EFE as a negative action-value using a policy $\phi=\phi(a_t|\mathbf{s}_t)$ (where $\mathbf{s}_t = \mathbf{o}_t$ as per our assumption earlier). The same policy $\phi$ is used for each future time step $\tau$, and the probability distribution over the first-step action is separated from $\phi$ resulting in a substitution distribution $q(a_t)$ for $\phi(a_t)$. Therefore, the one-step substituted EFE can be interpreted as the EFE of a policy of $(\phi(a_t), \phi(a_{t+1}),\dots,\phi(a_T)))$.

As in \cite{shin2021prior}, we simplify and approximate the search for optimal EFE values by adapting an estimation approach based on the Bellman equation, arriving at a Q-learning bootstrap scheme. Note that the Q-learning style framing of negative EFE estimation is referred to as G-learning. We utilize the AIF framework within the G-learning framing for the interception task and modify the framework to fit the interception task, see Figure~\ref{fig:AIF_arch}. Spatial variables, i.e. distance and speed, will serve as the inputs to our framework and, as mentioned before, an identity mapping is assumed to connect the observation directly to the state variables (allowing us to avoid having to learn additional parameterized encoder/decoder functions). As a result, the AIF agent we designed for this paper's experiments consists of two major components: a prior function and a multi-headed recognition neural model. 

Notably, our particular proposed recognition model works jointly as a function approximator of EFE values as well as a forward dynamics predictor. It takes in the current observation $\mathbf
{o}_t$ as input and then conducts, jointly, action selection and next-state prediction (as well as \textit{epistemic} value estimation). The selected action is executed and the resulting observation is returned by the environment. The prior function itself takes in as input the next observation $\mathbf{o}_{t+1}$, the consequence/result of the agent's currently selected action, and calculates the log likelihood of the preferred/prior distribution (set according to expert knowledge related to the problem), or the \textit{instrumental} term $R_{t,i}$. 
The difference between the outcome of the selected action $\mathbf{o}_{t+1}$ and its estimation $\hat{\mathbf{o}}_{t+1}$ (as per the generative transition component of our model) forms the \textit{epistemic} term $R_{t,e}$. The summation of the \textit{instrumental} and \textit{epistemic} terms forms the G-value (or negative EFE value) which is ultimately used to train/adapt the recognition model.
We explain each component in detail below.

\subsubsection{The Prior Preference Function and Prior Space}
\label{sec:prior_pref}

\begin{figure}[tb]
    \centering
    \includegraphics[width=0.45\textwidth]{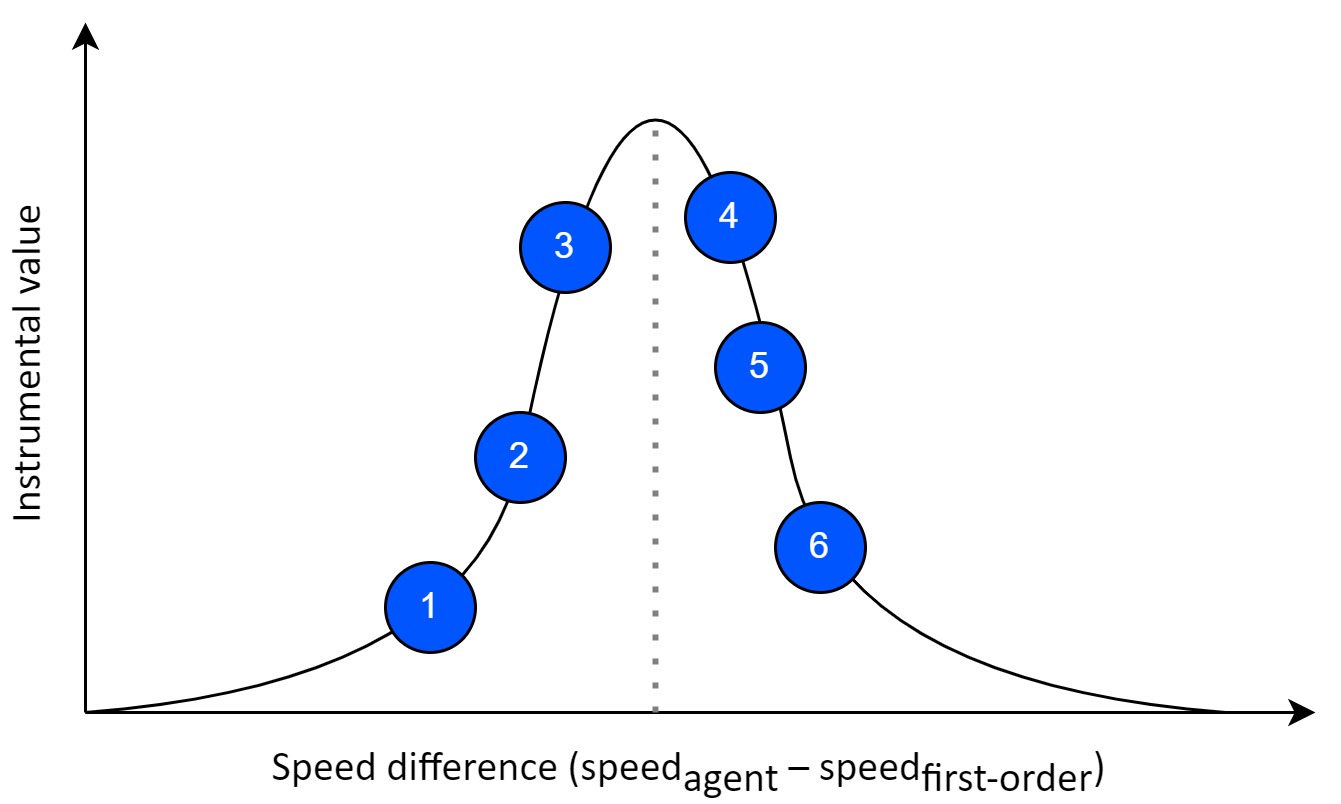}
    \caption{The prior preference specified in the prior space where each action corresponds to a different \textit{instrumental} value. Circles correspond to pedal positions to choose from.}
    \label{fig:prior_space}
    \vspace{-0.5cm}
\end{figure}

With the ability and freedom of designing a prior preference (or distribution over problem goal states) afforded by AIF, we integrate domain knowledge of the interception task into the design of a prior function. In essence, our designed prior function transforms the original observation vector $\mathbf{o}_t$ to a lower-dimensional space where a semantically meaningful variable is calculated (the prior space) and prior preference distribution is specified over this new variable -- in our case, this is set to be the \textit{speed difference}, as shown in Figure~\ref{fig:prior_space}. The \textit{speed difference} represents the difference between the agent's speed after taking the selected action and the speed required for successful interception, i.e. $\textit{speed difference} = speed_{agent}-speed_{required}$. Given the current observation, the required speed is calculated as the agent's distance to the interception point divided by the first-order target time-to-contact (TTC). We refer to this prior function as the \textit{first-order prior}. In our neural AIF framework, the \textit{instrumental} values are calculated given all possible actions (blue circles in Figure~\ref{fig:prior_space}) and a prior distribution over \textit{speed difference}. The smaller the \textit{speed difference} associated with a particular action, the higher the \textit{instrumental} value our function assigns. 

Note that the agent might not have enough time to adjust its speed later in the interception task if it only follows the guidance of this prior function without anticipating the likely future speed change of target, since this prior function only accounts/embodies first-order information.
To overcome this limitation, we investigated the effects of discounted long-term EFE value on the behavior of the agent in Section \ref{sec:discount_factor}.

\subsubsection{Recognition model}
\label{sec:recognition_model}

% Describe the Recognition model with EFE head and transition head
Our proposed recognition model embodies two key functionalities: EFE estimation and transition dynamics prediction, which are typically implemented as separate artificial neural networks (ANNs) in earlier AIF studies~\cite{shin2021prior} (in contrast, we found that, during preliminary experimentation, that a joint, fused architecture improved both the agent's overall generalization ability as well as its training stability). Concretely, we implement the recognition model as a multi-headed ANN with an EFE head and a transition head (see Figure~\ref{fig:AIF_arch}). The system takes in the current observation $\mathbf{o}_t$ and predicts: 
1) the EFE values for all possible actions, and 
2) a future observation at a distance $\mathbf{o}_{t+D}$ (in this work, we fix the temporal distance to be one step, i.e. $D=1$).
Within the recognition model, the current observation $\mathbf{o}_t$ is taken as input and a latent hidden activity vector $\mathbf{z}_t$ is produced, which is then provided to both output heads as input. The transition head $p(\mathbf{o}_{t+D}|\mathbf{z}_t)$ serves as a generative model (or a forward dynamics model) and the EFE head $G_\phi(\mathbf{z}_t, a)$ represents a variational density over the EFE values.
% justify why combing the transition model with the EFE model
As a result, EFE module and transition modules are wired together such that the prediction of the future observation $\mathbf{o}_{t+D}$ and the estimation of EFE values $\mathbf{G}_{t+D}$ are driven by the shared encoding from the topmost (hidden) layer of the recognition model.
This enables the sharing of underlying knowledge between the module selecting actions and the module predicting the outcome(s) of an action. Our intuition is that we humans tend to evaluate the ``value'' of an action by the consequences that it produces.

We next formally describe the dynamics of our recognition model, specifically its inference and learning processes.

\noindent
\textbf{Inference:}
In general, our agent is meant to produce an action conditioned on observations (or states) sampled from the environment at particular time-steps. Specifically, within any given $T$-step episode, our agent receives as input the observation $\mathbf{o}_t \in \mathcal{R}^{D \times 1}$, where $D$ is the dimensionality of the observation space $\mathbf{o}_t$ ($D = 4$ for the problem investigated in this study). The agent then produces a set of approximate free energy values, one for each action (similar in spirit to Q-values) as well as a prediction of the next observation that it is to receive from its environment (i.e., the perceptual consequence of its selected action).

Formally, in this work, the outputs described above are ultimately produced by a multi-output function $\mathbf{z}^3_a, \mathbf{z}^3_o = f_\Theta(\mathbf{o}_t)$, implemented as a multi-layer perceptron (MLP), where $\mathbf{z}^3_a$ contains estimated expected free energy values (one per discrete action) while $\mathbf{z}^3_o$ is the generative component's estimation of the next incoming observation $\mathbf{o}_{t+1}$. Note that we denote only outputting an action value set from this model as $\mathbf{z}^3_a = f^a_\Theta(\mathbf{o}_t)$ (using only the action output head) and only outputting an observation prediction as $\mathbf{z}^3_o = f^o_\Theta(\mathbf{o}_t)$ (using only the state prediction head). This MLP is parameterized by a set of synaptic weight matrices $\Theta = \{\mathbf{W}^1,\mathbf{W}^2,\mathbf{W}^3_a, \mathbf{W}^3_o \}$, that operates according to the following: 
\begin{align}
    \mathbf{z}^1 &= \phi_z(\mathbf{W}^1 \cdot \mathbf{\mathbf{z}^0}),\; \mathbf{z}^2 = \phi_z(\mathbf{W}^2 \cdot \mathbf{\mathbf{z}^1}) \\
    \mathbf{z}^3_a &= \phi_a(\mathbf{W}^3_a \cdot \mathbf{z}^2)), \; \mathbf{z}^3_o = \phi_a(\mathbf{W}^3_o \cdot \mathbf{z}^2))
\end{align}
where $\mathbf{z}^0 = \mathbf{o}_t$ (the input layer to our model is the observation at $t$). Note that a single discrete action is read out/chosen from our agent function's action output head as: $a = \argmax_a f^a_\Theta(\mathbf{o}_t)$. The linear rectifier $\phi_z(\mathbf{v}) = \text{max}(0, \mathbf{v})$ was chosen to be the activation function applied to the internal layers of our model while $\phi_a(\mathbf{v}) = \mathbf{v}$ (the identity) is the function specifically applied to the action neural activity layer $\mathbf{z}^3_a$ and $\phi_o(\mathbf{v}) = \mathbf{v}$ is the function applied to predicted observation layer neurons. 
Note that the first hidden layer $\mathbf{z}^1 \in \mathcal{R}^{J_1 \times 1}$ contains $J_1$ neurons and $\mathbf{z}^2 \in \mathcal{R}^{J_2 \times 1}$ contains $J_2$ neurons, respectively. The action output layer $\mathbf{z}^3_a \in \mathcal{R}^{A \times 1}$ contains $A$ neurons ($A = 6$ for the problem investigated in this study), one neuron per discrete action (out of $A$ total possible actions as defined by the environment/problem), while the observation prediction layer $\mathbf{z}^3_o \in \mathcal{R}^{D \times 1}$ contains $D=6$ neurons, making it the same dimensionality/shape as the observation space. 

\noindent
\textbf{Learning:}
\label{sec:learning}
While there are many possible ways to adjust the values inside of $\Theta$, we opted to design a cost function and calculate the gradients of this objective with respect to the synaptic weight matrices of our model for the sake of simulation speed. 
The cost function that we designed to train our full agent was multi-objective in nature and is defined in the following manner:
\begin{align}
    \mathcal{L}(\mathbf{o}_{t+1}, \mathbf{t}; \Theta) &= \mathcal{L}_a(\mathbf{o}_{t+1}; \Theta) + \mathcal{L}_o(\mathbf{t}_{t+1}; \Theta) \\
    \mathcal{L}_a(\mathbf{t}; \Theta) &= \frac{1}{2\sigma^2_a}||\mathbf{t} - \mathbf{z}^3_a||^2_2 \\
    \mathcal{L}_o(\mathbf{o}_{t+1}; \Theta) &= \frac{1}{2\sigma^2_o}||\mathbf{o}_{t+1} - \mathbf{z}^3_o||^2_2
\end{align}
where the target value for the action output head is calculated as $t_j = r_j + \gamma \max_a f^a_\Theta(\mathbf{o}_t)$
while the target action vector is computed as $\mathbf{t}_j = t_j \mathbf{a}_j + (1 - \mathbf{a}_j) \otimes f^a_\Theta(\mathbf{o}_t)$.
In the above set of equations, we see that the MLP model's weights are adjusted so as to minimize the linear combination of two terms, the cost associated with the difference between a target vector $\mathbf{t}$, which contains the bootstrap-estimated of the EFE values, and the agent's original estimate $\mathbf{z}^3_a$ as well as the cost associated with how far off the agent's prediction/expectation $\mathbf{z}^3_o$ of its environment is from the actual observation $\mathbf{o}_{t+1}$. In this study, the standard deviation coefficients associated with both output layers are set to one, i.e., $\sigma_a = \sigma_o = 1$ (highlighting that we assume unit variance for our model's free energy estimates and its environmental state predictions -- note that a dynamic variance could be modeled by adding an additional output head responsible for computing the aleatoric uncertainty associated with $\mathbf{o}_{t+1}$).

Updating the parameters $\Theta$ of the neural system then consists of computing the gradient $\frac{\partial \mathcal{L}(\mathbf{o}_{t+1}, \mathbf{t};\Theta)}{\partial \Theta}$ using reverse-mode differentiation and adjusting their values using a method such as stochastic gradient descent or variants, e.g., Adam~\cite{kingma2014adam}, RMSprop~\cite{tieleman2012lecture}. Specifically, at each time step of any simulated episode, our agent first stores the current transition of the form $(\mathbf{o}_t,\mathbf{a}_t,r_t,\mathbf{o}_{t+1})$ into an episodic memory replay buffer and then immediately calculates  $\frac{\partial \mathcal{L}(\mathbf{o}_{t+1}, \mathbf{t};\Theta)}{\partial \Theta}$ from a batch of observation/transition data (uniformly) sampled from the replay buffer, which stores up to $10^5$ transitions. 
We will demonstrate the benefit of this design empirically in the results section.

\section{Results}
\label{sec:results}

\subsection{Hypotheses for Interception Strategies}
\label{sec:hypotheses}

Given the fact that the target changes its speed during a trial in our interception task, the agent / human subject could gain advantage by anticipating the target speed change prior to the change of target speed. To select an optimal action early within the trail, the agent needs to take into consideration the initial target speed in the current trial and make adjustments based on the experience acquired from previous trials.
So, the question becomes: how does the agent adapt its behavior on the basis of current trials’ observation of target speed / distance from the interception point and the learned statistics across trials?

\subsection{Experimental Setup}
We implemented the interception task as an environment in Python based on the OpenAI gym~\cite{brockman2016openai} library. This integration provides the full functionality and usability of the gym environment, which means that the environment can work / be used with any RL algorithm and is made accessible to the machine learning community as well. Our AIF agents and baseline algorithm DQN are implemented with the Tensorflow2~\cite{tensorflow2015-whitepaper} library. Experimental data and code will be made publicly available upon acceptance. % INSERT GITHUB LINK HERE...

\subsection{Task Performance}
\label{sec:performance}

\begin{figure}[tb]
    \centering
    \includegraphics[width=0.9\textwidth]{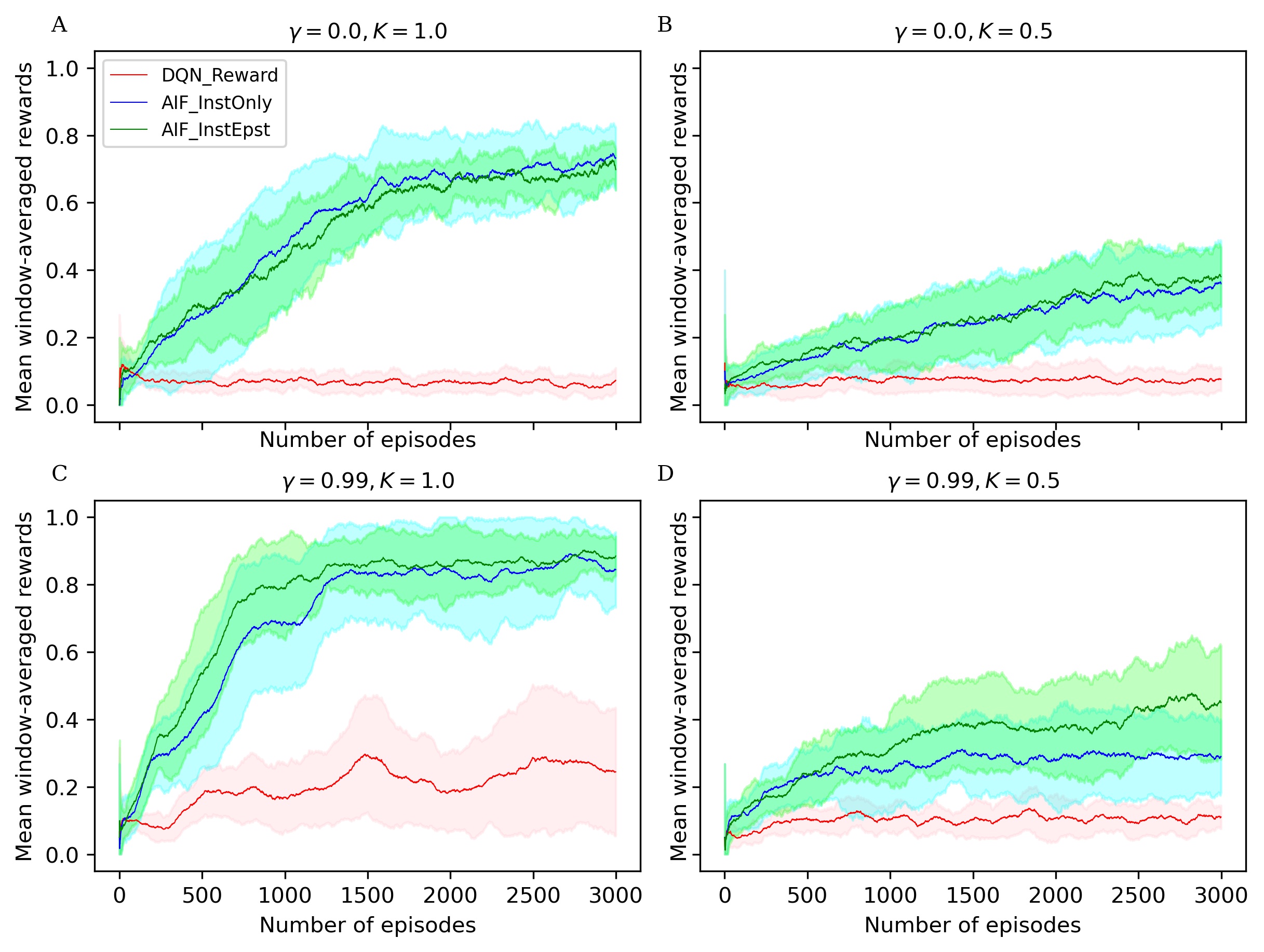}
    \caption{Window-averaged reward measurements of agent performance on the interception task. $DQN\_Reward$ represents a DQN agent that utilized the sparse reward signal and $\epsilon-greedy$ exploration; $AIF\_InstOnly$ represents our AIF agent with only \textit{instrumental} component which is defined by the prior function; $AIF\_InstEpst$ represents an AIF agent that consists of both \textit{instrumental} and \textit{epistemic} components. Discount factor is denoted by $\gamma$, pedal lag coefficient is denoted by $\textit{K}$.}
    \label{fig:perf_compare}
\end{figure}

We compare AIF agents with and without the \textit{epistemic} component and a baseline algorithm, i.e. a deep-Q network (DQN). Experiments are conducted for $20$ trials where each trial contains $3000$ episodes. The task performance of agents is shown as curves plotting window-averaged rewards (with a window size of $100$ episodes) in Figure~\ref{fig:perf_compare}, where the solid line depicts the mean value across trials and the shaded area represents standard deviation.
We conducted a set of experiments where the discount factor $\gamma$ of the models and the pedal lag coefficient $\textit{K}$ were varied (note that, in AIF and RL research, $\gamma$ is typically fixed to a value between $0.9$ and $1$ to enable the model to account for long term returns). In order to compare the performance of our agents to that of human subjects, we apply the original pedal lag coefficient in one set of our experiments (specifically shown in Figure~\ref{fig:perf_compare}C).

Observe that our AIF agents are able to reach around a $90\%$ success rate stably with very low variance. This beats human performance with $47\%$ (std=11.31) on average and $54.9\%$ in the final block of experiments reported in~\cite{diaz2009intercepting}.
The baseline DQN agent, which learns from the problem's sparse reward signal at the end of each episode, yields an average success rate of $22\%$ at test time. Similarly, the AIF agent with both \textit{instrumental} and \textit{epistemic} components achieves a $90\%$ mean success rate.

Note that the DQN agent is outperformed by the AIF agents trained with our customized prior preference function by a large margin. This reveals that the flexibility of injecting prior knowledge is crucial for solving complex tasks more efficiently and validates our motivation of applying AIF to cognitive tasks. 
In our preliminary experiments, we tested an AIF agent which consists of an EFE network and a transition network separately. This AIF agent is out-performed by the AIF agent with recognition model in terms of windowed mean rewards and stability. Furthermore, the AIF agent with recognition model has lower model complexity.
Specifically, AIF agent with recognition model has only $66.8\%$ of the parameter counts of that of AIF agent with separate models. This supports our intuition that combining the EFE model with the transition model yields an overall better model agent.

Interestingly, the AIF agent with only \textit{instrumental} component was able to nearly reach the same level of performance as the full AIF agent. However, success rate of this agent exhibited a larger variance than the full AIF agent. Based on comparison between agents with and without \textit{epistemic} component, we argue that \textit{epistemic} component serves, at least in the context of the interception task we investigate, as a regularizer for the AIF models, providing improved robustness. Since we apply experience replay and bootstrapping to train the AIF models, it is possible that a local minimum is reached in the optimization process because the replay buffer is filled up with samples which come from the same subspace as the state space. Therefore, with the help of \textit{epistemic} component, the agent is encouraged to explore the environment more often and adjusts its prediction of future observations such that it has a higher chance of escaping poorer local optima.
Our proposed AIF agent reaches a plateau in performance after about $1000$ episodes and stabilizes more after $1500$ episodes. Note that, in contrast, human subjects were able to perform the task at an average success rate after $9$ episodes of initial practice~\cite{diaz2009intercepting}.

\subsection{Anticipatory Behavior of AIF Agents}
\label{sec:discount_factor}

\begin{figure}[!hb]
    \centering
    \includegraphics[width=0.9\textwidth]{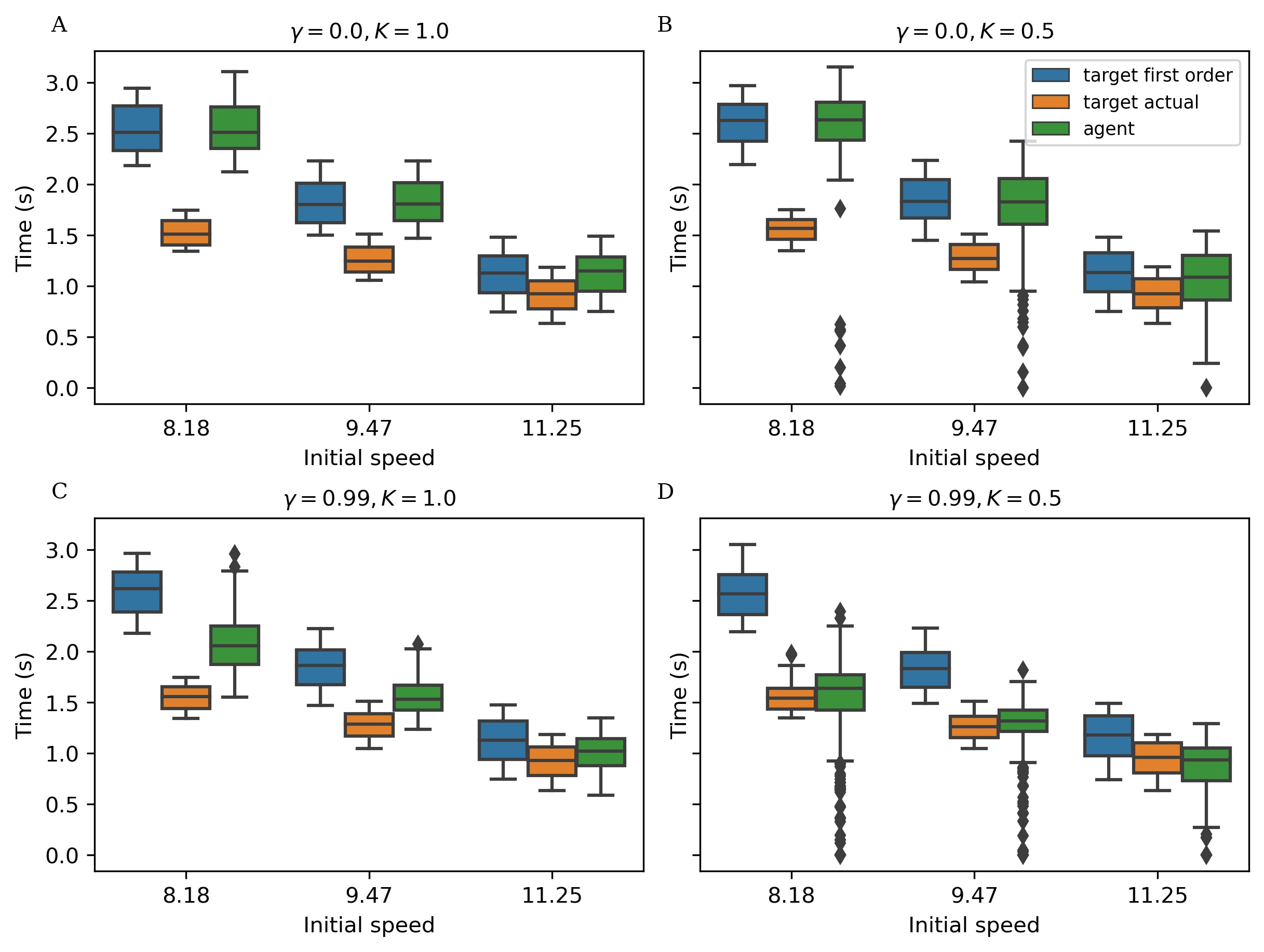}
    \caption{TTC values taken at the onset of target's speed change. In each subplot, the target's first-order TTC, the target's actual mean TTC, and the agent's TTC are shown in different colors, with data grouped by target initial speed. The discount factor is denoted by $\gamma$ while the pedal lag coefficient is denoted by $\textit{K}$.}
    \label{fig:ttc_compare}
\end{figure}

Do the AIF agents exhibit a similar capacity for anticipatory behavior as humans do?
To answer this question and to compare the strategy used by our AIF agents to that of human subjects, we record the Time-To-Contact (TTC) from trained AIF agents at the onset of the target's speed change in each episode. We then calculate, at the same time: 
1) the target's TTC using first-order information, and 
2) target's TTC with the assumption that the target would change its speed at the most likely time and reach an averaged final speed. Finally, we compose these three types of TTC data grouped by target initial speed into a single boxplot in Figure~\ref{fig:ttc_compare}.

In our experimental analysis, we found that the discount factor $\gamma$ plays a big role in forming different behavior patterns within AIF agents. All variants of AIF agents were trained with the \textit{instrumental} value computed using our first-order prior function. Intuitively, the agent's behavior should conform to a reactive agent who uses only the first-order information and acts to match its own TTC to the target's first-order TTC, just like what has been observed in Figure~\ref{fig:ttc_compare}A (please see that the green box is nearly identical to the blue box under all target initial conditions). The AIF agent depicted in Figure~\ref{fig:ttc_compare}A is set to use a discount factor of $0$, which means that the agent only seeks to maximize its immediate reward without considering the long-term impact of the action(s) that it selects. Such an agent converges to a reactive behavior. However, when we increase the discount factor to $0.99$ (which is a common practice in RL literature), the AIF agent starts to behave more interestingly. In Figure~\ref{fig:ttc_compare}C, the agent's TTC (green box) lies in between target's first-order TTC (blue box) and target's actual mean TTC (orange box), which suggests that the AIF agent tends to move faster than a pure-reactive, first-order agent would in the early phase of interception. In other words, the agent tends to anticipate the likely target speed change in the future and adjusts its action selection policy. This behavioral pattern can be explained as exploiting the benefits provided by estimating long-term accumulated \textit{instrumental} reward signal (when the discount factor value is increased). Given a higher discount factor, in this case $\gamma=0.99$, the AIF agent estimates the summation of \textit{instrumental} values from its current (time) step in the task until the end of the interception using discounting. This leads to an agent who seeks to maximize long-term benefits in terms of reaching the goal when selecting actions.

\subsection{Effect of Vehicle Dynamics on Agent Behavior}
\label{sec:pedal_lag}

To test how anticipatory behavior is affected when simple reactive behavior is no longer sufficient, we increased the inertia on the agent's vehicle by changing the pedal lag coefficient $\textit{K}$. Given the same discount factor $\gamma=0.99$, we compare two different pedal lag coefficients $\textit{K}=1.0$ in Figure~\ref{fig:ttc_compare}C and $\textit{K}=0.5$ in Figure~\ref{fig:ttc_compare}D, where lower $\textit{K}$ indicates less responsive vehicle dynamics.
With the same discount factor, the AIF agent performing the task under a lower pedal lag coefficient in Figure~\ref{fig:ttc_compare}D has a lower success rate in intercepting the target. This is due to the fact that the agent's ability to manipulate its own speed is limited, therefore there is less room left for error. However, the AIF agent in this condition yields TTC values that are closer to the target's actual mean TTC. Note that, when the target initial speed is $11.25\ m/s$ (Figure~\ref{fig:ttc_compare}D), the median of agent's TTC value is actually smaller than target's actual mean TTC. This supports our hypothesis that purely reactive behavior is not sufficient for successful interception and anticipatory behavior is emergent when the vehicle becomes less responsive.

\section{Discussion}

Variations of an AIF agent were trained to manipulate the speed of movement so as to intercept a target moving across the ground plane, and eventually across the agent's linear path of travel. On each trial, the target would change in speed on most trials to a value that was selected from a Gaussian distribution of final speeds. The results demonstrate that the AIF framework is able to model both on-line visual and anticipatory control strategies in an interception task, as was previously demonstrated by humans performing the same task \cite{diaz2009intercepting}. The agent's anticipatory behavior aimed to maximize the cumulative expected free energy in the duration that follows action selection. Variation of the agent's discount factor modified the length of this duration. At lower discount factors, the agent behaved in a reactive manner throughout the approach, consistent with the constant bearing angle strategy of interception. At higher values, actions that were selected before the predictable change in speed took into account the most likely change in target speed that would occur later in the trial.  Anticipatory behavior was also influenced by the agent's capabilities for action.This anticipatory behavior was most apparent when the pedal lag coefficient was set to lower values, which had the effect of changing the agent's movement dynamics so that purely reactive control was insufficient for interception behavior.

\begin{figure}[tb]
    \centering
    \includegraphics[width=0.5\textwidth]{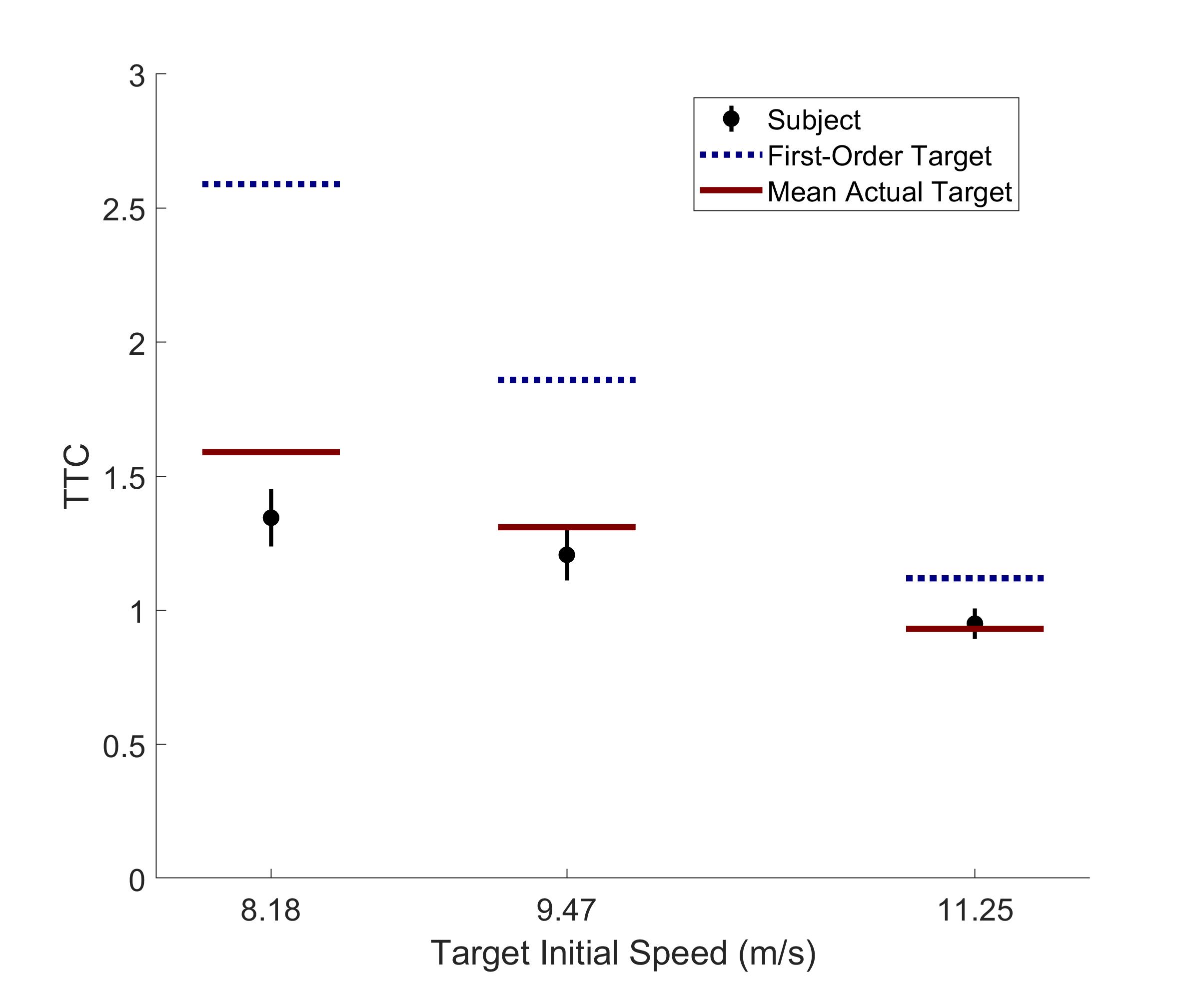}
    \caption{Human subject data from Exp 1. of~\cite{diaz2009intercepting}. TTCs were taken at the onset of target's speed change. Dotted line represents the mean of target's first-order TTC, solid line represents the mean of target's actual TTC, black disk represents the mean of subject's TTC with a bar indicating $95\%$ confidence interval of the mean.}
    \label{fig:human_data}
\end{figure}

% There are some ways in which the "equivalent model" differed from humans
%%% GD:  Paragraph revised for clarity 10/7 . Original commented out below.
Despite the agent's demonstration of qualitatively human-like prediction, careful comparison of the agent's behavior to the human performance and learning rates demonstrated in \cite{diaz2009intercepting} reveals notable differences. Analysis of participant behavior in the fourth and final block of Experiment 1 in \cite{diaz2009intercepting} reveals that subject TTC at the onset of the target's change in speed was well matched to the most likely time and magnitude of the target's likely change in speed (i.e., the mean actual target TTC in Figure~\ref{fig:human_data}). In contrast, the AIF agent with an equivalent pedal lag ($\textit{K}$= 1.0; i.e. the \textit{matched} agent)  demonstrated only partial matching of its TTC to the likely change in target speed (the target's mean actual TTC in Figure \ref{fig:ttc_compare}C). Although one might attribute this to under-training of the agent, it is notable that the agent achieved a hit rate exceeding 80\% by the end of training, while human participants in the original study consistently improved in performance until reaching 55\% hit rate at the end of the experiment. 

% Review of representation head, instrumental, and epistemic reward
% EFE is reward based learning, consistent with dopaminergic system.
% Trans. head is strong model based.
To better understand the potential causes of these differences between agent and human performance, it is helpful to consider how the agent's mechanism for anticipation differs from that of humans. The agent chooses actions on the basis of a weighted combination of reward-based reinforcement (instrumental reward) and short model-based prediction (epistemic reward), both of which are computed within the two-headed recognition model. Instrumental reward is computed in the EFE head, which is responsible for selecting the action (i.e., pedal position) that it estimates would produce the greatest expected free energy later in the agent's approach. The estimate of EFE associated with each pedal position does not involve an explicit process of model-based prediction, but is learned retrospectively, through the use of an experience replay buffer. Following action selection, visual feedback provides an indication of the cumulative EFE over the duration of the replay buffer. The values of EFE within this buffer are weighted by their temporal distance from the selected action in accordance with the parameter of discount factor. This is similar to both reward-based learning and is often compared to the dopaminergic reward system in humans \cite{haruno_neural_2004,lee_neural_2012,holroyd_neural_2002,momennejad_successor_2017}. The epistemic component of the EFE reward signal is thought to drive exploration towards uncertain world states, and it relies on predictions made in the transition head.  This component of the model relies on the hidden states provided by the shared neural layers in the recognition model and predicts an observation at next time step $\hat{\textbf{o}}_{t+1}$. The estimated observation at next time step is then compared to the ground truth observation $\textbf{o}_{t+1}$ and the difference between them generates the epistemic signal $R_{t,e}$. The role of the transition head is in many ways consistent with a ``strong model-based'' form of prediction \cite{ZHAO2015190}, whereby predictive behaviors are planned on the basis of an internal model of world states and dynamics that facilitate continuous extrapolation. In summary, whereas the EFE head is consistent with reward based learning, the transition head is consistent with relatively short-term model based prediction.

%  Brief review of human anticipation. This may be redundant with text in the introduction.
How does this account of anticipation demonstrated by our agent compare with what we know about anticipation in humans? As discussed in the introduction, empirical data on the quality of model-based prediction suggests that it degrades sufficiently quickly that it cannot explain behaviors of the sort demonstrated here, by our agent, or by the humans in \cite{diaz2009intercepting}. In contrast, a common theory in motor control and learning relies upon a comparison of a very short-term prediction (e.g. milliseconds) of self-generated action with immediate sensory feedback \cite{blakemore1998central,wolpert1995internal,hoist1950reafferenzprinzip,wade1994hermann}.  However, this similarity is weakened by the observation that, in the context of motor-learning, short-term prediction is thought to rely upon access to an efferent copy of the motor signal used to generate the action. For this reason, it is problematic that the AIF agent is predicting both its own future state ($x_s, v_s$) and the future state of the target ($x_t, v_t$), for which there is no efferent copy or analogous information concerning movement dynamics. Although research on eye movements has revealed evidence for the short-term prediction of future object position and trajectory \cite{diaz2013saccades, diaz_memory_2013,ferrera_internally_2010}, it remains unclear whether these behaviors are the result of predictive models of object dynamics or representation-minimal heuristics.

%%% Here are some ways in which we could modify our model to be more consistent with hybrid control.
% Optical variables that vary in reliability 
Another possible contribution to the observed differences between agent and human performance is the perceptual input. When considering potential causes for the difference between agent and human anticipatory behavior, it is notable that the agent relies upon an observation vector defined by agent's and target's position and velocity measured in meters, and meters per second, respectively. However, in the natural context, these spatial variables must be recovered or estimated on the basis of perceptual sources of information, such as the rate of global optic flow due to translation over the ground plane, the exocentric direction of the target, the instantaneous angular size of the target, or the looming rate of the target during the agent's approach. It is possible that by depriving the agent of these optical variables, we are also depriving the agent of opportunities to exploit task-relevant relationships between the agent and environment, such as the bearing angle. It is also notable that some perceptual variables may provide redundant information about a particular spatial variable (e.g. both change in bearing angle and rate of change in angular size may be informative about an objects approach speed).  However, redundant variables will differ in reliability by virtue of sensory thresholds and resolutions. For these reasons, a more complete and comprehensive model of human visually guided action and anticipation would take as input potential sources of information and learn to weight them according to context-dependent reliability and variability.

%% Motor delay
Another potential contributor to differences between human and agent performance is the notable lack of visuo-motor delays within the agent's architecture.  In contrast, human visuo-motor delay has been estimated to be on the order of 100-200 ms between the arrival of new visual information and the modification or execution of an action \cite{nijhawan_visual_2008, le_runigo_visuo-motor_2010}. Because uncompensated delays would have devastating consequences on human visual and motor control, they are often cited as evidence that humans must have some form of predictive mechanism that acts in compensation \cite{wolpert1995internal}. Future attempts to make this model's anticipatory behavior more human-like in nature may do so by imposing similar length delay between the agent's choice of motor plan on the basis of the observed world-state and the time that this motor plan is executed~\cite{walsh2009learning}. 

\subsection{Conclusion}
We present a novel generalized active inference framework (AIF) model for studying online visually guided locomotion using an interception task where a moving target changes its speeds in a semi-predictable manner. In order to drive the agent towards the goal more effectively, we devised a problem-specific prior function, improving the agent's computational efficiency and interpretability. Notably, we found that our proposed AIF agent exhibits better task performance when compared to a commonly used RL agent, i.e. the deep-Q network (DQN). The full AIF agent, containing both \textit{instrumental} and \textit{epistemic} components, exhibited slightly better task performance and lower variance compared to the AIF agent with only an \textit{instrumental} component. Furthermore, we demonstrated behavioral differences among our full AIF agents given different discount factor $\gamma$ values as well as levels of the agent's action-to-speed responsiveness. Finally, we analyzed the anticipatory behavior demonstrated by our agent and examined the differences between the agent's behavior and human behavior. 
While our results are promising, future work should address the following limitations - first, inputs to our agent are defined in a simplified vector space whereas sensory inputs to the humans that actually perform the interception task are visual in nature (i.e., the model should work directly with unstructured sensory data such as pixel values). We remark that a vision-based approach could facilitate the extraction of additional information and features that are useful for solving the interception task more reliably. Second, our simulations do not account for visuo-motor delays inherent to the human visual and motor systems, and that might be modeled using techniques like delayed Markov decision process formulations~\cite{walsh2009learning, firoiu2018human}.

\bibliographystyle{acm}
\bibliography{refs}

\end{document}